# Universality of the Intermediate Phase and High-Temperature Superconductivity in $C_{60}$

J. C. Phillips

*Bell Laboratories, Lucent Technologies\*, Murray Hill, N. J. 07974-0636*

With the discovery of $T_c$'s over 100K, continuously biased hole-doped semiconductive $C_{60}^{+y}$ has joined the family of high-temperature superconductors (HTSC) previously dominated by the ceramic cuprates. The phase diagrams of these two quite different materials are remarkably similar, despite the absence from undoped $C_{60}^{+y}$ of magnetic order. This supports percolative models of HTSC based on strong electron-phonon interactions. These models explain observed fine structure in the $C_{60}^{+y}$ phase diagrams, and furthermore suggest novel paths to still higher $T_c$'s.

## Introduction

When $C_{60}$ was first discovered, it was expected that the physical properties of materials formed from this large, rigid molecule, bound together by weak interactions, would be radically different from those of previously known materials. However, the recent discovery of high-temperature superconductivity (HTSC) in $C_{60}^{+y}$ interfaces doped by gate bias has revealed phase diagrams[1,2] remarkably similar to those of (by now conventional) cuprate HTSC. There have been a great many theories of HTSC in the cuprates, and almost all of them now appear to be inapplicable to $C_{60}^{+y}$. Thus one is faced with a major conceptual choice. One can assume that the similarity is accidental, and develop separate and independent theories for HTSC in the two materials, thus creating as many "flavors" of theory as there are classes of materials[3]. Alternatively one can assume that there are deep reasons for the similarity, and that these deep reasons can be understood theoretically. This reduction of assumptions (economy of means, recommended by both Aristotle and William of Ockham, around 1300) will narrow the theoretical possibilities by a very large factor, of order perhaps $10^2 - 10^3$.

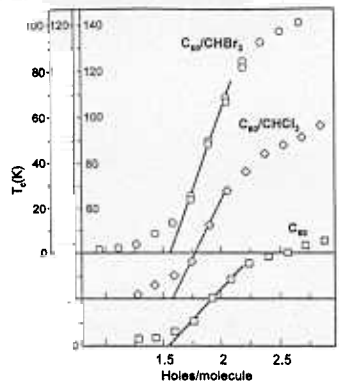

Figure 2

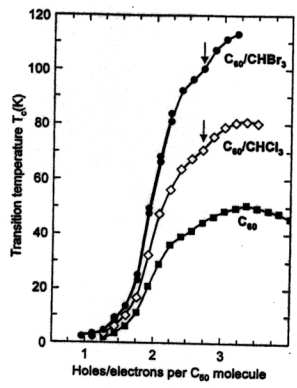

Figure 3

This paper is based on Ockham's razor, and to support this approach it presents a very detailed analysis of phase diagrams of $C_{60}^{+y}$ in the framework of a coherent percolative topological approach previously used to describe the cuprates[4,5]. In this approach many microscopic details are different between the two systems, but the overall mechanisms responsible for HTSC are quite similar. This does not mean that the microscopic details are irrelevant; indeed, they are necessary chemical factors that must be optimized in each class of material to maximize $T_c$, for example. It does mean, however, that these microscopic factors in both cases, when optimized, are part of a common, and indeed generic or universal, mechanism that generates HTSC.

The phase diagrams of the cuprates, often modeled as parabolic, but which are actually more nearly trapezoidal, have been analyzed in detail[5,6] using the quantum percolation model. Percolation, especially when it is coherent, produces physical properties that differ drastically from those obtained within the effective medium approximation (EMA). The latter is, of course, much simpler, and in the face of complex behavior, it appears to be the first choice of almost everyone. (From this one should not conclude, as many have done, that the EMA is correct; for example, there are many different EMA models that have been constructed to explain various properties of ceramic HTSC, and nearly all of these models are mutually inconsistent.) However, one can see from many examples, both classical molecular and quantum electronic, that the EMA often fails[5-7] to describe connectivity (transport) phase transitions based on *attractive* interactions in *disordered* systems. The reason for that failure is that fluctuations, which are either neglected entirely or treated as small and random by the EMA, become *strongly correlated*, as enhanced connectivity lowers the free energy, even in the normal state, and even in classical molecular systems with no phase coherence. These strong correlations, which cannot be treated by perturbation theory, lead to dimensional collapse and to the formation of one-dimensional filaments, providing that there is enough (glassy) disorder in the system. The disordered elements, such as dopants, should be mobile enough that the self-organized, zigzag filamentary dopant band is kinetically accessible at reasonable annealing temperatures. The correlations need not be, and almost certainly are not, primarily dynamical fluctuations at low temperatures.

In addition to phase diagrams the quantum percolation model explains many, many properties of the cuprates that are anomalous in the EMA. These include linearities in T of the *normal state* resistivity and Hall number[8], mysterious correlations between vibronic peaks in *neutron* and *infrared* spectra (never discussed together in experimental papers), the broken symmetry of planar and c-axis polarized infrared spectra[9], conflicts between tunneling spectra that show an isotropic energy gap and angle-resolved photoemission spectra that suggest d-wave anisotropy[10], and doping-dependent medium-range correlations in atomic-resolution scanning tunneling microscope coupled with rapid short-range fluctuations[10,11]. In other words, the quantum percolation model is not just a toy model that explains a few aspects of one experiment, it is a general platform for understanding all the microscopic properties of cuprate HTSC.

The recent data on high-temperature superconductivity (HTSC) in $C_{60}^{+y}$ induced by gate doping in a field-effect transistor configuration have so far been restricted primarily to measurements of the temperature-dependent resistivity, which includes $T_c(y)$. The most striking aspect of these data is that the composition $|y_{max}| = 3$ at which $T_c$ is maximized is the same for $y > 0$ as for $y < 0$. It has generally been assumed that the reason $|y_{max}| = 3$ for $y < 0$ is that the lowest conduction band level (lowest unoccupied molecular orbital, LUMO) is orbitally three-fold degenerate, or 6-fold degenerate including spin. Thus $|y_{max}| = 3$ corresponds to a half-filled band. However, if this reasoning were correct, then for $y > 0$ one would have $y_{max} = 5$, instead of the observed value $y_{max} = 3$, as the highest valence band level (highest occupied molecular orbital, HOMO) is five-fold orbitally degenerate. This puzzle reveals fundamental inadequacies in continuum theories of the phase diagrams of $C_{60}^{+y}$. The theory of quantum percolation, developed over the last decade for the cuprates, readily resolves this fundamental puzzle without introduction of new assumptions, thus agreeing with Ockham's razor.

In addition to the major puzzle of electron-hole symmetry in $|y_{max}| = 3$, the observed phase diagrams contain many subtle features, including breaks in slope. If these data are considered in isolation, then the many breaks in slope of $T_c(y)$ that are discussed here might be difficult to explain. However, the similarities to the cuprate phase diagrams are so pervasive that one is well justified in transferring much of the reasoning that has been so successful in the cuprates to the $C_{60}^{+y}$ data. Care must first be taken to include the specific effects associated with the confinement of the dopant space charge to a disordered surface layer in contact with the gate oxide. These effects, fortunately, turn out to be quite tractable.

## 2. Percolative Transport in a Depletion Space Charge Layer

In field-effect transistor geometries, which are sometimes described as d = two-dimensional, substantial disorder from ionized impurities in the oxide and/or interfacial roughness gives rise to metal-insulator transitions (MIT) at low temperatures. To date the most successful description of this transition employs a model that combines global classical percolation with local quantum tunneling across saddle point barriers[12]. Global classical percolation is demonstrated by the lowest temperature data of conductance vs. density, which follows the s = 1.3 power law predicted by numerical simulations in two dimensions. Quantum point contact tunneling sets the overall scale of the conductance as $e^2/h$, and the temperature dependence is consistent with a resistor network with an exponential distribution of resistances. Within each minimum, or "puddle", the current is carried incoherently until a perpendicular magnetic field is applied. One can then recognize a greatly broadened first Landau level associated with the average puddle size.

The difference between MIT in d = 2 and d = 3 is that in the latter case it is possible to have coherent current flow along the entire current path, either in a randomly distributed impurity band[4,7], as T → 0, or in a self-organized impurity band, as in well-annealed cuprates[5]. It is this coherence that gives rise to linearities in T of the normal state resistivity and Hall number[8]. In the fullerene case, the resistivity is quadratic in T, in another words, the normal state behaves as a Fermi liquid because the resistivity is associated with current flow across the puddles. In the cuprates there is a close correlation between the normal-state anomalies and optimal values of $T_c$. Thus phenomenologically one would not expect that the superconductive phase diagrams of the cuprates and $C_{60}^{+y}$ would be similar. However, they are, and the filamentary platform enables one to see easily why this is so.

## 3. Percolative Thresholds and Nanoscale Immiscibility in Cuprates and $C_{60}^{+y}$

The characteristic signature of a miscibility gap is the appearance of spinodal tie lines across the immiscibility dome. At the boundaries of the dome, where the spinodal connects to the pure phases, one expects to find breaks in slope of observed quantities, which should be linear, or nearly so, along the spinodal line. These breaks in slope have recently been identified[6] for $La_{2-x}Sr_xCuO_4$ and correlated with thresholds in the superconductive filling factor, as measured by the Meissner effect and the magnitude of the specific heat jump at $T_c$. Many similar breaks are observed[1] in $T_c(y)$ for $C_{60}^{+y}$. For the reader's convenience we show first the phase diagrams for $La_{2-x}Sr_xCuO_4$ in Fig. 1.

Two curves are shown in Fig.1. The lower curve is the filling factor $f(x)$. In any percolative theory on some scale $f(x)$ becomes equivalent to the usual percolative site or bond occupation probability p. On shorter scales $f(x)$ represents an average over short-range fluctuations. In the cuprates nanodomains with diameters ~ 3 nm occur because of ferroelastic stress relief[9]. Because of orthorhombic nanodomain correlations, these fluctuations may extend to somewhat larger lengths, say ~ 10 nm. In any case, as a result of these static fluctuations, the percolation of superconductivity is strongly coupled to internal lattice strains. This broadens $T_c(x)$ compared to $f(x)$, and it introduces spinodal phase separation at the two phase transitions that determine the boundaries of the intermediate superconductive phase. (In complex structures like YBCO, additional spinodal regions may occur[6], for example, between the 60K and 90K plateaus in $T_c$.) In Fig. 1 the correspondence between the two transitions in f(x) and the two spinodal regions in $T_c(x)$ is self-evident.

In the spinodal regions $T_c(x)$ is nearly linear, as one would expect from an effective medium model. This may indicate that the scale of orthorhombic correlations is indeed much larger than the nanodomain length scale itself, as indicated above. Other microscopic properties, such as the isotope shift[13], are also observed to be linear in the spinodal regions, and further linearities may well be recognized as the nature of these regions become more widely known.

The most complete fullerite phase diagram obtained so far is for unexpanded $C_{60}^{+y}$, as shown in Fig. 2 for two samples. The better sample shows slightly larger $dT_c(y)/dy$ in the first spinodal region. There are several breaks in slope of $T_c(y)$. The dashed spinodal lines have been drawn to emphasize the similarities between Figs. 1 and 2. The two spinodal regions are expected to be dissimilar. The first one, centered near y = 2, is derived from a continuous percolative transition from disconnected filaments to connected ones. The second region is associated with the transition from the filamentary metallic phase to the Fermi liquid (normal metal, not superconductive). The second transition is first order, because the filaments are destroyed abruptly. There is a larger "tail" where $T_c(y) > 0$ in the overdoped phase, because it is more difficult to achieve equilibrium near a first-order transition, than near a continuous (essentially displacive) transition. These features have been observed not only in $La_{2-x}Sr_xCuO_4$, Fig. 1, and other cuprates, but also in ideally simple impurity bands[4], such as neutron transmutation doped Ge, and even in the reversibility window of network glass transitions[7].

## 4. Metal-Insulator Transition in $C_{60}^{+y}$

It has been clear for some time that in general there is a close connection between HTSC and the metal-insulator transition (MIT), but it appears that this connection has been addressed in a concrete way only within the framework of filamentary percolation theory[4,5]. However, there is no evidence for such a correlation in the resistivity data[1,2] on $C_{60}^{+y}$, and this suggests that there are fundamental differences between cuprate and $C_{60}$ phase diagrams and HTSC. Near and above the doping onset of HTSC, the resistivity increases smoothly with temperature in the cuprates, with $1 < d\log\rho(T)/d\log T < 2$. Doesn't this mean that the two kinds of HTSC are fundamentally different?

"Obviously" the answer to this question is "yes". However, in HTSC things are seldom what they seem to be. To begin, even in the cuprates there are wide variations between the kinds of correlations one finds between the onset of HTSC and the MIT. Most of the correlations are rather broad, and are obscured by pseudogap effects. The most striking case is $YBa_2Cu_3O_x$ at the metal-insulator transition at $x = 6.4$, where superconductivity also starts and the host crystal structure changes from tetragonal to orthorhombic. In samples annealed under pressure above 240K, $dT_c/dP$ is by far the largest here[6] among all the HTSC. Topologically it seems very likely that this behavior reflects the appearance of long chain segments in the $CuO_{1-x}$ plane, which complete internanodomain ballistic filamentary paths.

To explain the complete absence of an MIT correlated with the threshold for HTSC, one should go to the opposite limit, which would suggest that in $C_{60}^{+y}$, most of the normal state transport is diffusive. In particular, in the normal state it appears that much of the resistivity arises from librational disruption of increased overlap between Jahn-Teller distorted spheroidal $C_{60}$ molecules. We return to these distortions below.

The $T_c(y)$ data on $C_{60}^{+y}$, as expanded by inclusion of dipole orientable halogenated methane $[CH(Cl,Br)_3]$ are shown in Fig. 3. Most exciting is the large increase in $T_c(y)$, which was expected by analogy with the effect on $T_c$ of expansion of alkali-doped $C_{60}^{-w}$. Equally intriguing theoretically are the linear first spinodal regions associated with the MIT. As shown in Fig. 4, when these are extrapolated to $T_c(y), = 0$, they all have a nearly common intercept $y = y_0 = 1.55(2)$.

In general one would not have expected that expansion would increase $T_c$. For example[14], in $La_{2-x}Sr_xCuO_4$, the reverse occurs: expansion by growth of thin films on substrates with larger (smaller) lattice constants actually decreases (increases) $T_c$. When $T_c(x_{max})$ increases, $x_{max}$ changes from 0.16 to 0.10, and a very narrow first spinodal region is observed with an MIT at x = 0.045. However, it appears that in this case there have been large changes in the nanodomain wall filling factor and dopant segregation factor with change of substrate[15]. Unlike the perovskites, the fullerites are not ferroelastic, so nanodomains are not expected to form, and the simpler behavior that is actually observed becomes possible.

Returning to Fig. 3, one notices another major break in slope near y = 2.7. The feature may be absent in $C_{60}^{+y}$, but it is pronounced with addition of $CHCl_3$, and even more pronounced with further increase of $T_c$ by addition of $CHBr_3$. It is difficult to explain this $T_c$-correlated feature simply in terms of lattice expansion. Suppose, however, that because of their attractive dipolar interactions the added $CHCl_3$, $CHBr_3$ molecules orient to form head-to-tail chains, as has occurred in some numerical simulations of dipolar systems at low densities[16]. These curvilinear chains reinforce the film (much as steel rods are used to reinforce concrete), and lead to larger local lattice constant expansion. The regions adjacent to the chains would have larger local $T_c$'s, and when y is large enough for these regions to become metallic on their own, one would have an enhanced $T_c$, much as observed. As the effect is topological, it would be common to both Cl and Br methanated films at nearly the same value of y. Note that, while the actual jump in $T_c$ associated with this feature is small, its overall effect on $T_c$ may be much larger, as it is present as a large background interaction even when the chains are not fully connected. Note also that mean-field expansion effects scale with volume expansion $\Delta V$ as $\Delta V^s$ with s = 1, but that filamentary effects scale with s = 1/3 (or 1/2 in a surface layer). For small $\Delta V$ the filamentary effects, if present, should be much larger than the mean field effects.

The oxide/$C_{60}^{+y}$ interface is rough, not planar. The highest valence levels of $C_{60}$ are five-fold degenerate. As holes are added, there is a Jahn-Teller splitting, and the first hole occupies an orbital chain that has been illustrated recently[17]. Then one can expect the partially ionized C atoms in this chain to expand, changing the spheroidal $C_{60}$ into a prolate ellipsoid. The orientation of this ellipsoid will be determined by Coulombic interactions between the space charge depletion layer and ionized impurities in the oxide. One would expect that most of the ellipsoids would orient with the most highly polarizable prolate axis normal to the interface. This would reduce intermolecular overlap in the $C_{60}^{+y}$ plane, giving rise to an insulating phase. By increasing y above 1, holes will begin to occupy and cause Jahn-Teller splittings of a second orbital, causing a second axis in the plane to expand, making larger planar overlap and planar metallic conduction possible. This enhanced connectivity should occur at $y - 1 = 0.50$, as the ellipsoidal topology of the glassy surface layer is essentially that of the square lattice (regardless of the geometry of the bulk $C_{60}$ film ), and the mean-field value[18] for percolation in a square lattice is 0.5. The value $y = y_0 = 1.50$ just calculated is in excellent agreement with the extrapolated experimental value $y_0 = 1.55(2)$ quoted above for the *filamentary superconductive* MIT. As $y_0$ is entirely topological, and involves no coupling constants, it is independent of $T_c$ and applies equally well to all three films, again in good agreement with experiment.

Enhanced connectivity associated with planar Jahn-Teller distortions depends on "freezing" these distortions into orientations that maximize intermolecular overlap. Some of this freezing may be produced by internal fields associated with interfacial roughness, and some of it can be caused by energy gains associated with filamentary superconductivity - local Cooper pair formation. In any case, this connectivity can be reduced by thermal librational fluctuations of the $C_{60}^{+y}$ spheroids. Thus it seems likely that it is this combination of factors that explains the absence of anomalous structure in $\rho(T,y)$ – it is erased by librational motion ($\omega \sim 10\text{-}100$ cm$^{-1}$) in the normal state[19,20]. At the same time, the frequency of these motions is low enough that local Cooper pair formation, involving energies of order $T_c$, can freeze planar Jahn-Teller distortions of adjacent $C_{60}^{+y}$ spheroids into orientations that maximize intermolecular overlap leading to filamentary formation. Such freezing is, of course, a first-order effect, and it would be the microscopic mechanism associated with the first spinodal shown in Figs. 2 and 4. Note that there is no natural place in continuum or Fermi liquid dynamical theories for such static first-order effects.

The success of the topological filamentary model in calculating $y_0$ naturally leads to the question, can the model calculate $y_{max}$, the concentration that maximizes $T_c$? The observed value of $y_{max}$ is about 3.3, where the second spinodal begins (Fig. 2). In the absence of nanoscale phase separation $y_{max}$ could be as large as 4. The calculation of spinodal boundaries is considered to be extremely difficult, even for simple classical liquids, and it will not be attempted here. One can remark that qualitatively above $y = 3$ the intermolecular overlap will become nearly isotropic, which is unfavorable for filamentary formation, but favorable for a Fermi liquid. Just as in the cuprates, the Fermi liquid phase is not superconductive, because the Coulomb interactions overwhelm the attractive e-p interactions at low carrier densities in an isotropic geometry. The ratio $y_{max}/y_0$ is ~ 2, much as in Si:P impurity bands[4].

A striking feature of the experimental data[1] is that the values of $y_{max}$ are nearly the same for electron (whether chemically with alkalies, or by biasing) and hole doping. This is not what one would have expected in the EMA from the differing densities of states and orbital degeneracies [x] of the conduction [3] and valence bands [5], but then there is no correlation between EMA predictions based on undistorted Fermi energy densities of states $N(E_F,y)$ and experimental $T_c(y)$ at any level. A more significant question is how the electron-hole similarities are to be explained within the present model.

First, one can note that in the FET configuration (biased doping), the *ellipsoidally distorted topological* model is *independent of carrier sign*, and so the observed electron-hole similarities are a very successful intrinsic feature of the model. Thus the model has only to explain why the chemical doping with alkalies yields the same results as biased doping without alkalies. It would appear that the reported crystal structures of "$A_3C_{60}$" are incompatible with glassy filamentary formation, but it must be remembered that diffraction measurements, based on coherent scattering, on partially ordered (mixed glassy and crystalline regions) samples, see only the ordered regions. There is no doubt that the interfacial depletion layer studied with the FET configuration is highly disordered. This leads directly to the conclusion that, even in $A_3C_{60}$, the observed superconductivity could well be associated with percolation in disordered regions even of single-crystal samples.

## 5. Some Microscopic Comparisons

The microscopic theory of $T_c$ in terms of electron-phonon interactions in $C_{60}$ and its bulk (especially alkali-doped, y = -3) derivatives is well developed, for instance, in terms of semi-empirical tight-binding models[21], or muffin-tin methods[22], which yield similar results, including a value[22] for y = +2. The e-p coupling constant $\lambda = NV$ depends on the interball N and intraball V's. Because of the $\pi$ character of Fermi energy states the modes with the largest V's are primarily low-frequency buckling modes and high-frequency tangential modes, with smaller contributions from radial modes. In spite of the large uncertainties in $\lambda$, the values predicted for $T_c$ are in reasonably good agreement with experiment, except that in the alkali derivatives (y = -3) the predicted increase in $\lambda$ with lattice expansion is about 50% larger than that implied by the observed increase in $T_c$. For $C_{60}$ (+2) the predicted value[20] of $T_c$ = 45K is in good agreement with the observed value[1] of 30K.

A recent development[23] is the discovery of superconductivity in $C_{70}^{-y}$ with a similar narrow phase diagram as in $C_{60}^{-y}$, with $y_{max}(70) = 4$ compared to $y_{max}(60) = 3$, and $T_c^{max}(70) = 7K$, compared to $T_c^{max}(60) = 11K$. The larger magnitude of $y_{max}(70)$ may be the result of the pre-formed prolate ellipsoidal structure[24,25]. Given the general trend[26] of $T_c^{max}$ towards smaller values with larger numbers of $\pi$-bonded C atoms, the smaller value of $T_c^{max}(70)$ is reasonable. On the basis of this result, it is suggested that one possible way to reach larger values of $T_c^{max}$ is to replace $C_{60}$ by $C_{36}$.

There is a second way to increase $T_c^{max}$ that might be easier. As noted above, the presence of $CH(Cl,Br)_3$ may have increased $T_c^{max}$ in two ways: the obvious way, through overall lattice expansion and band narrowing, and a less obvious way, by guiding the formation of filaments in more strongly expanded glassy regions adjacent to dipolar $CH(Cl,Br)_3$ chains or chain segments. In the d = 2 depletion space-charge geometry one can picture $T_c^{max}$ being reduced not by long-range Coulomb interactions, but by short-range Coulomb blockade in a weak link of a percolative path. The scale[12] of Coulomb blockade effects is $e^2/h$. It has been reported[27] that these effects can be reduced by a factor of 25 by replacing (100) Si substrates by vicinal substrates (tilt angle ~ $10^o$). The vicinal steps may orient filaments and enhance $T_c^{max}$ in much the same way that $T_c^{max}$ is enhanced from the 60K plateau to the 90K plateau in YBCO by ordering in the $CuO_{1-x}$ plane.

## 6. Conclusions

Our discussion has shown that there are both similarities and differences between the transport properties of the cuprates and $C_{60}^{+y}$. In the normal state the interesting anomalies observed in the cuprates are absent from $C_{60}^{+y}$, which behaves as a normal Fermi liquid, probably because of strong librational scattering by Jahn-Teller distorted spheroids. (In effect, the intermolecular band width is modulated by librational motion.) This motion is frozen in the hole-doped strongly superconductive phase, which leads to filamentary formation, as in the cuprates. Thus the intermediate superconductive phase is similar in the two kinds of materials, and there are very good topological reasons for the similarities of the superconductive phase diagrams of $La_{2-x}Sr_xCuO_4$ and $C_{60}^{+y}$.

One of the characteristic features of topological theories is that they sometimes are able to predict *magic numbers* for compositions at which phase transitions occur, as in the especially favorable case of network glasses[7] with the magic fraction 12/5. In the cuprates this has not proved to be the case, because of the tendency of dopants to segregate in nanodomain walls[15]. In the fullerenes the large $C_n$ molecules (n = 60, 70, etc.) naturally partition space discretely, much as the nanodomains do in the cuprates. When the dopant level and the Jahn-Teller distortions are determined continuously by the internal field, the situation is similar to that in network glass alloys where the connectivity is varied continuously by changing the average coordination number. Then it may be possible to observe *magic numbers.* Indeed, the data show magic compositions $y_0 = 1.5$ and $y_{max} = 3.0$ and 4.0. Such integral and half-integral magic compositions never occur in continuum or Fermi liquid theories. Of course, the data presented so far are sparse, but as the database grows it will be interesting to see whether or not the appearance of these magic numbers is accidental.

In principle it is still possible, but highly implausible, that the microscopic origin of HTSC in these two classes of materials could be different[3]. However, the strong similarities identified and explained here suggest that it will be very difficult to invent alternative mechanisms. Because traditional electron-phonon interactions are so successful for the $C_{60}$ family, this leads unavoidably to the conclusion that e-p interactions must also be regarded as the origin of HTSC in the cuprates as well. Glassy disorder is an essential cofactor in both families.

*Postscript.* The universality of the phase diagram has been confirmed in a recent study of single-layer $CaCuO_{2+z}$ films in the FET geometry[28]. The features identified here for $La_{2-x}Sr_xCuO_4$ and $C_{60}^{+y}$ are also observed for this "infinite layer" hole-doped HTSC. In the notation of Fig. 1, the corresponding values are $x_{11} = 0.06$, $x_{12} = 0.10$, and $x_{21} = 0.20$, and, as in Fig. 4, $y_0 = 0.05$. Although the doping level is controlled by the applied field, the authors note that apical oxygen (z) may well play an important part (as a hidden dopant) in the superconductivity. This is suggested by high-pressure experiments by others whom they cite. Here again apical oxygen may well be present because of compression due to epitaxial effects. It is an oversimplification to suppose that the apical oxygen functions merely as a charge reservoir; this naïve picture fails badly[9] for YBCO.

*Retired. jcphillips8@home.com

## Figure Captions

Fig. 1. Percolative interpretation[6] of data on the filling factor f(x), measured by Meissner volume and specific heat jump $\Delta C_p(x)$, and $T_c(x)$ in $La_{2-x}Sr_xCuO_4$. *Two* phase transitions in f(x) are reflected in *two* spinodal immiscibility domes in $T_c(x)$, indicated by the dotted lines. These two transitions are the boundaries of the intermediate HTSC phase.

Fig. 2. $T_c(y)$ for $C_{60}^{+y}$ from Ref. 1.

Fig. 3. $T_c(y)$ for $C_{60}^{+y}$ and as expanded by inclusion of $CHCl_3$ and $CHBr_3$, from Ref. 2.

Fig. 4. The data from Fig. 3, offset to show the extrapolation of $T_c(y)$ to $T_c = 0$. The common intercept occurs at y = 1.55(2), in good agreement with the topological value of 1.50.

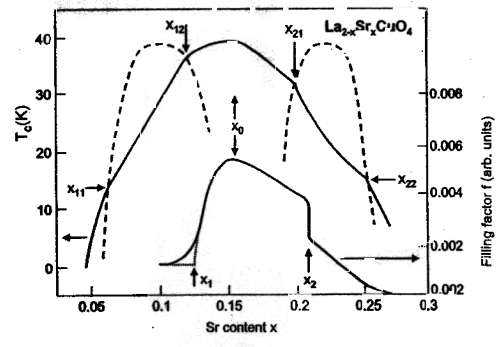

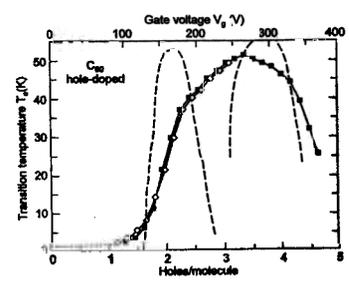

Figure 2

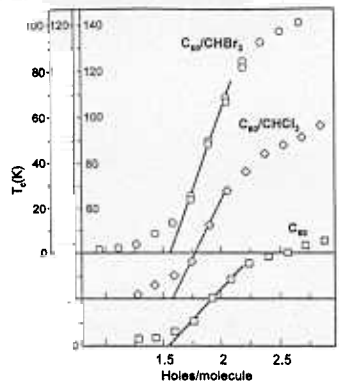

Figure 2

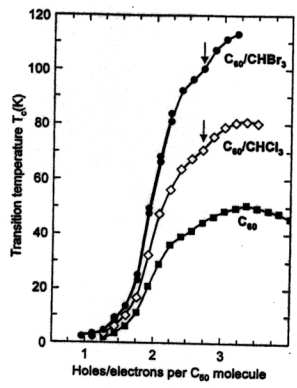

Figure 3